\documentclass[11pt,a4paper]{article}
\usepackage{amssymb}
\usepackage{amsmath}
\usepackage{amsfonts}
\usepackage{bbm}
\usepackage{amsthm}
\usepackage{mathrsfs}
\usepackage{hyperref}
\usepackage{color}
\usepackage[margin=2.41cm]{geometry}
\usepackage[all,cmtip]{xy}
\usepackage[utf8]{inputenc}
\usepackage{graphicx}
\usepackage{varwidth}
\usepackage{comment}
\usepackage{enumitem}

\usepackage{upgreek}
\usepackage{rotating}

\usepackage{tikz}
\usetikzlibrary{shapes.geometric}

\usepackage{todonotes}
\usepackage[most]{tcolorbox}

\definecolor{darkred}{rgb}{0.8,0.1,0.1}
\hypersetup{
	colorlinks=true,         % false: boxed links; true: colored links,false is default
	linkcolor=darkred,
	citecolor=blue,
}

\theoremstyle{plain}
\newtheorem{theo}{Theorem}[section]

\newtheorem{propo}[theo]{Proposition}

\theoremstyle{definition}
\newtheorem{defi}[theo]{Definition}

\newtheorem{exx}[theo]{Example}
\newenvironment{ex}
{\pushQED{\qed}\exx}
{\popQED\endexx}

\newtheorem{remm}[theo]{Remark}
\newenvironment{rem}
{\pushQED{\qed}\remm}
{\popQED\endremm}

\newtheorem{openn}[theo]{Open Problem}
\newenvironment{open}
{\pushQED{\qed}\openn}
{\popQED\endopenn}

\numberwithin{equation}{section}

\def\nn{\nonumber}

\def\bbK{\mathbb{K}}
\def\bbR{\mathbb{R}}
\def\bbC{\mathbb{C}}

\def\bbZ{\mathbb{Z}}

\def\id{\mathrm{id}}

\def\dd{\mathrm{d}}

\def\1{I}
\def\oone{\mathbbm{1}}

\def\Loc{\mathbf{Loc}}

\def\Set{\mathbf{Set}}
\def\Alg{\mathbf{Alg}}

\def\Vec{\mathbf{Vec}}
\def\Ch{\mathbf{Ch}}

\def\CC{\mathbf{C}}
\def\DD{\mathbf{D}}

\def\MM{\mathbf{M}}
\def\NN{\mathbf{N}}
\def\TT{\mathbf{T}}
\def\Cat{\mathbf{Cat}}

\def\Op{\mathbf{Op}}
\def\dgAQFT{\mathbf{dgAQFT}}
\def\AQFT{\mathbf{AQFT}}
\def\tPFA{\mathbf{tPFA}}
\def\dgCat{\mathbf{dgCat}}
\def\Pr{\mathbf{Pr}}

\def\AAA{\mathfrak{A}}

\def\BBB{\mathfrak{B}}

\def\FFF{\mathfrak{F}}

\def\AAAA{\boldsymbol{\mathfrak{A}}}

\def\O{\mathcal{O}}
\def\P{\mathcal{P}}

\DeclareMathOperator*{\bigboxtimes}{\text{\raisebox{-0.5ex}{\scalebox{1.4}{$\boxtimes$}}}}

\def\sk{\vspace{2mm}}

\makeatletter
\let\@fnsymbol\@alph
\makeatother

\title{%
Operads, homotopy theory and higher categories \\
in algebraic quantum field theory
}

\author{%
Marco Benini$^{1,2,a}$\ and\ Alexander Schenkel$^{3,b}$\vspace{4mm}\\
{\small ${}^1$ Dipartimento di Matematica, Universit\`a di Genova,}\\
{\small Via Dodecaneso 35, 16146 Genova, Italy.}\vspace{2mm}\\
{\small ${}^2$ INFN, Sezione di Genova,}\\
{\small Via Dodecaneso 33, 16146 Genova, Italy.}\vspace{2mm}\\
{\small ${}^3$ School of Mathematical Sciences, University of Nottingham,}\\
{\small University Park, Nottingham NG7 2RD, United Kingdom.}\vspace{4mm}\\
{\small 
Email:  ${}^a$~\href{mailto:marco.benini@unige.it}{\texttt{marco.benini@unige.it}}~,~~${}^b$~\href{mailto:alexander.schenkel@nottingham.ac.uk}{\texttt{alexander.schenkel@nottingham.ac.uk}}\vspace{2mm}
}
}

\date{April 2023}

\begin{document}

\maketitle

\begin{abstract}
\noindent This chapter provides a non-technical overview and motivation for the recent interactions between algebraic quantum field theory (AQFT) and rather abstract mathematical disciplines such as operads, model categories and higher categories.
\end{abstract}
\vspace{-1mm}

\vspace{-2mm}

\vspace{-1mm}

\tableofcontents

%%%%%%%%%%%%%%%%%%%%%%%%%%%%%%%%%%%%%%%%%%%%%%%%
%%%%%%%%%%%%%%%%%%%%%%%%%%%%%%%%%%%%%%%%%%%%%%%%

\section{\label{sec:intro}Introduction}
\textit{Algebraic quantum field theory} (in short, AQFT)
is a well-developed and time-honored mathematical framework in which one can
define and study quantum field theories (QFTs) on the Minkowski spacetime \cite{HaagKastler} or, 
more generally, on globally hyperbolic Lorentzian manifolds \cite{BFV}. AQFT has already had many
interesting and fruitful interactions with a broad range of mathematical disciplines, 
including operator algebras, functional analysis, Lorentzian geometry
and also category theory. We refer the reader to the 
book \cite{AQFTbook} for an overview, as well as
to the other chapters of this volume that
cover different aspects of AQFT.
\sk

As the title suggests, the focus of this chapter is on the relatively
recent interactions between AQFT and rather abstract areas of pure mathematics
such as operad theory, homotopy theory (implemented through model categories) and higher category theory. 
The main goal is to illustrate the motivations 
for such interactions and the achievements they lead to, 
which can be concisely, yet incompletely, summarized as follows: 
\begin{itemize}
\item Via operad theory, one can identify and describe
the fundamental algebraic structures underlying AQFT. This leads to
a plethora of universal constructions that are, among other things, useful
to (i)~understand deeply the time-slice axiom (which, informally speaking, 
encodes the concept of time-evolution in AQFT), leading to classification 
results for AQFTs in low dimensions, (ii)~construct new AQFTs out 
of old ones, with examples given by local-to-global extensions 
(refining Fredenhagen's universal algebra \cite{Fredenhagen}) or the extraction of the
chiral observables of a $2$-dimensional conformal AQFT (refining Rehren's work \cite{Rehren}),
and (iii)~compare AQFT with other axiomatizations of QFT, such as factorization algebras {\`a} la Costello
and Gwilliam \cite{CG1,CG2}.

\item Via abstract homotopy theory, see e.g.\ \cite{Hovey,Hirschhorn}, one can develop
model categories whose objects are, for instance, cochain complex
valued AQFTs and whose weak equivalences are spacetime-wise quasi-isomorphisms.
This is the natural habitat in which examples of gauge-theoretic AQFTs that 
are constructed in terms of the BRST/BV formalism \cite{FR1,FR2} live.
Furthermore, abstract homotopy theory is also needed to develop a gauge-theoretic generalization
of retarded and advanced Green's operators, which in particular 
opens up new avenues to extend the well-established free (i.e.\ non-interacting) AQFT constructions
from \cite{Bar,BGP} to gauge-theoretic and also higher gauge-theoretic examples.
Finally, when combined with operad theory, insights can be obtained about higher homotopical phenomena 
in the underlying algebraic structure of AQFTs (or a lack thereof).

\item Via higher category theory, one can introduce a vastly generalized concept of 
AQFT that does not rely on the existence of suitable function algebras on the classical
moduli spaces of fields. Instead of algebras, such AQFTs 
assign linear categories or dg-categories, with examples given by quantizations 
(in the sense of derived algebraic geometry \cite{Toen,CPTVV})
of quasi-coherent modules on the derived moduli stacks of fields. 
In contrast to the usual BRST/BV formalism, this approach
is non-perturbative in the sense that it carries information about the gauge group
and not only its Lie algebra. Unfortunately, constructing examples
of such (dg-)category valued AQFTs is rather involved and presently only simple toy-models are understood.
\end{itemize}

This chapter will substantiate and explain in more detail the items above.
Our presentation is intentionally informal, which will unavoidably 
lead to omissions. To compensate for these shortcomings, 
we will provide wherever appropriate references in which precise statements 
and the relevant technical details can be found. We do not expect readers
to be familiar with operads, homotopy theory or higher categories, even though 
some background will be helpful to more easily follow the narrative of this chapter.
A detailed introduction to these subjects is beyond the scope of this chapter.
However, we shall explain the main ideas informally 
and where useful pictorially.

%%%%%%%%%%%%%%%%%%%%%%%%%%%%%%%%%%%%%%%%%%%%%%%%
%%%%%%%%%%%%%%%%%%%%%%%%%%%%%%%%%%%%%%%%%%%%%%%%

\section{\label{sec:operad}Operads}

\paragraph{The basic idea:} An \textit{operad} is an object 
that captures $n$-to-$1$ operations together with their behavior under composition.
One of the simplest and most illustrative examples is given by the unital associative
operad $\mathsf{uAs}$. This operad is generated by a $0$-to-$1$
operation (the unit) and a $2$-to-$1$ operation (the multiplication), which 
we shall visualize as follows
\begin{subequations}\label{eqn:uAs}
\begin{flalign}
\parbox{2cm}{\begin{tikzpicture}[cir/.style={circle,draw=black,inner sep=0pt,minimum size=2mm},
        poin/.style={circle, inner sep=0pt,minimum size=0mm},scale=0.8, every node/.style={scale=0.8}]
\node[poin] (Uout)  at (2,1) {};
\node[poin] (Uin)   at (2,0) {};
\draw[thick] (Uin) -- (Uout);
\draw[thick,fill=white] (Uin) circle (0.6mm);
\node[poin] (Mout)  at (4,1) {};
\node[poin] (Min1)  at (3.7,0) {};
\node[poin] (Min2)  at (4.3,0) {};
\node[poin] (V)     at (4,0.5) {};
\draw[thick] (Min1) -- (V);
\draw[thick] (Min2) -- (V);
\draw[thick] (V) -- (Mout);
\end{tikzpicture}}\quad.
\end{flalign}
These operations should compose in a unital and associative fashion, i.e.\
\begin{flalign}
\parbox{6cm}{
\begin{tikzpicture}[cir/.style={circle,draw=black,inner sep=0pt,minimum size=2mm},
        poin/.style={circle, inner sep=0pt,minimum size=0mm},scale=0.8, every node/.style={scale=0.8}]
%
%
%\node[poin] (label1) [label=center:{$l_t,\, r_t:$}] at (-1.5,0.5) {};
\node[poin] (Mout)  at (0,1) {};
\node[poin] (Min1)  at (-0.3,0) {};
\node[poin] (Min2)  at (0.3,0) {};
\node[poin] (V)     at (0,0.5) {};
\draw[thick] (Min1) -- (V);
\draw[thick] (Min2) -- (V);
\draw[thick] (V) -- (Mout);
\draw[thick,fill=white] (Min1) circle (0.6mm);
\node[poin] (EqM) at (0.75,0.5) {$=$};
\node[poin] (MMout)  at (1.25,1) {};
\node[poin] (MMin)   at (1.25,0) {};
\draw[thick] (MMin) -- (MMout);
\node[poin] (EqMM) at (1.5,0.5) {$~=$};
\node[poin] (MMMout)  at (2.25,1) {};
\node[poin] (MMMin1)  at (1.95,0) {};
\node[poin] (MMMin2)  at (2.55,0) {};
\node[poin] (VVV)     at (2.25,0.5) {};
\draw[thick] (MMMin1) -- (VVV);
\draw[thick] (MMMin2) -- (VVV);
\draw[thick] (VVV) -- (MMMout);
\draw[thick,fill=white] (MMMin2) circle (0.6mm);
\node[poin] (Nout)    at (4.5,1.25) {};
\node[poin] (Nin1)    at (4.2,0.25) {};
\node[poin] (Nin2)    at (7.3,0) {};
\node[poin] (NV)      at (4.5,0.75) {};
\node[poin] (Ninin1)  at (3.9,-0.25) {};
\node[poin] (Ninin2)  at (4.5,-0.25) {};
\node[poin] (Ninin3)  at (5.1,-0.25) {};
\draw[thick] (Nin1) -- (NV);
\draw[thick] (Ninin3) -- (NV);
\draw[thick] (Ninin1) -- (Nin1);
\draw[thick] (Ninin2) -- (Nin1);
\draw[thick] (NV) -- (Nout);
\node[poin] (EqN) at (5.5,0.5) {$=$};
\node[poin] (NNout)    at (6.5,1.25) {};
\node[poin] (NNin2)    at (6.8,0.25) {};
\node[poin] (NNV)      at (6.5,0.75) {};
\node[poin] (NNinin1)  at (5.9,-0.25) {};
\node[poin] (NNinin2)  at (6.5,-0.25) {};
\node[poin] (NNinin3)  at (7.1,-0.25) {};
\draw[thick] (NNinin1) -- (NNV);
\draw[thick] (NNin2) -- (NNV);
\draw[thick] (NNV) -- (NNout);
\draw[thick] (NNinin2) -- (NNin2);
\draw[thick] (NNinin3) -- (NNin2);
\end{tikzpicture}
}\quad,
\end{flalign} 
\end{subequations}
where the $1$-to-$1$ operation in the unitality conditions is the identity operation.
All operads appearing in this chapter will be symmetric operads, i.e.\ there is
an action of the permutation group $\Sigma_n$ on the set of $n$-to-$1$ operations
that is compatible in a suitable sense with compositions, see e.g.\ \cite{Yau} for the details.
\sk

Operads can be generalized in many ways, and the most relevant generalization for us is to decorate
the inputs and outputs with colors. The resulting concept is that of \textit{colored (symmetric) operads},
which for brevity we shall simply call operads.
The most basic example here is a category $\CC$: The objects $c\in \CC$ determine the colors
and the morphisms $f:c\to c^\prime$ can be thought of as $1$-to-$1$ operations. 
Note that there are no $n$-to-$1$ operations, for $n \neq 1$, when regarding a category $\CC$ as an operad.
\sk

Given any (colored symmetric) operad $\O$, one can form its category
$\Alg_\O(\TT)$ of algebras with values in a symmetric monoidal category $(\TT,\otimes,I,\gamma)$, 
e.g.\ that of vector spaces $\Vec_\bbK$ over a field $\bbK$.
An algebra over the operad $\O$ should be thought of as a representation
of the abstract operations described by $\O$ in terms of morphisms in $\TT$. More precisely, an algebra
$\AAA\in \Alg_\O(\TT)$ consists of a family of objects $\AAA(c)\in\TT$, for all colors $c\in\O$, 
together with a family of $\TT$-morphisms
\begin{flalign}
\AAA(\phi)\,:\, \bigotimes_{i=1}^n \AAA(c_i)\,\longrightarrow\,\AAA(c^\prime)\quad,
\end{flalign}
for all $n$-to-$1$ operations $\phi : (c_1,\dots,c_n)\to c^\prime$ in $\O$,
that is compatible with compositions, identities and permutation actions.
Not surprisingly, an algebra $\AAA\in \Alg_{\mathsf{uAs}}(\TT)$ 
over the unital associative operad is a unital associative algebra in $\TT$,
with unit $I\to \AAA$ and 
multiplication $\AAA\otimes \AAA\to \AAA$ respectively given by representing 
the $0$-to-$1$ and $2$-to-$1$ operations from \eqref{eqn:uAs}.
Note that an algebra $\AAA\in \Alg_{\CC}(\TT)$ over the operad determined by a category $\CC$
is a functor $\AAA:\CC\to \TT$.

\paragraph{The AQFT operad:} AQFTs are typically defined on the category
$\Loc_m$ of connected $m$-dimensional oriented and time-oriented globally hyperbolic Lorentzian manifolds,
with morphisms given by orientation and time-orientation preserving causal isometric embeddings, see
e.g.\ \cite{AQFTbook}.
This category comes with distinguished classes of 
(tuples of) morphisms that capture important Lorentzian geometric features. For the definition
of AQFTs and hence of the AQFT operad, the following two classes are essential:
\begin{itemize}
\item[(i)] A pair of morphisms $f_1:M_1\to N \leftarrow M_2:f_2$ to a common target
is called \textit{causally disjoint} if their images can not
be connected by a causal curve, i.e.\ $J_N(f_1(M_1))\cap f_2(M_2) =\emptyset$
with $J_N(S):= J^+_N(S) \cup J^-_N(S)\subseteq N$ the union of
the causal future and past of a subset $S\subseteq N$. We often
denote causally disjoint morphisms by $f_1\perp f_2$.

\item[(ii)] A morphism $f:M\to N$ in $\Loc_m$ is called a \textit{Cauchy morphism}
if its image $f(M)\subseteq N$ contains a Cauchy surface of $N$. We denote
the class of Cauchy morphisms by $W\subseteq \mathrm{Mor}\,\Loc_m$.
\end{itemize}

The usual definition of an $m$-dimensional AQFT is in terms of a functor
$\AAA : \Loc_m\to \Alg_{\mathsf{uAs}}(\TT)$ 
to a category of unital associative algebras that satisfies the Einstein 
causality and time-slice axioms. This can be reformulated in a more elegant and
conceptual way as follows (see \cite[Theorem 2.9]{2AQFT} for the details): 
An AQFT is given by the assignment
\begin{subequations}\label{eqn:AQFTpictures}
\begin{flalign}\label{eqn:AQFTpictures1}
\parbox{8cm}{
\begin{tikzpicture}[scale=0.5]
\filldraw[fill=gray!20, draw=black] (0,0) -- (1.8,1.8) -- (3.6,0) -- (1.8,-1.8) -- (0,0);
\draw (3.1,1.3) node {\footnotesize{$M$}};
\draw[very thick, |->] (5,0) -- (8,0) node[midway,above] {{\footnotesize{\text{AQFT}}}};
\draw (13,0) node {$\mathfrak{A}(M) \in \Alg_{\mathsf{uAs}}(\TT)$};
\end{tikzpicture}}
\end{flalign}
of a unital associative algebra to each spacetime $M\in \Loc_m$, together with the assignment
\begin{flalign}\label{eqn:AQFTpictures2}
\parbox{8cm}{
\begin{tikzpicture}[scale=0.5]
\filldraw[fill=gray!20, draw=black] (0,-4) -- (1.8,-2.2) -- (3.6,-4) -- (1.8,-5.8) -- (0,-4);
\draw (3.1,-2.7) node {\footnotesize{$N$}};
\draw[thick, dotted] (1,-3.2) -- (2.6,-4.8);
\draw[thick, dotted] (1,-4.8) -- (2.6,-3.2);
\draw[fill=gray!70, draw=black] (0.2,-4) -- (0.7,-3.5) -- (1.2,-4) -- (0.7,-4.5) -- (0.2,-4);
\draw (0.7,-4) node {\tiny{$M_1$}};
\draw (1.8,-4) node {\text{{\bf $\cdots$}}};
\draw[fill=gray!70, draw=black] (2.4,-4) -- (2.9,-3.5) -- (3.4,-4) -- (2.9,-4.5) -- (2.4,-4);
\draw (2.9,-4) node {\tiny{$M_n$}};
\draw[very thick, |->] (5,-4) -- (8,-4) node[midway,above] {{\footnotesize{\text{AQFT}}}};
\draw (13,-4) node {$\bigotimes\limits_{i=1}^n \AAA(M_i) ~\longrightarrow ~\AAA(N)$};
\end{tikzpicture}}
\end{flalign}
of a unital associative algebra morphism to each tuple $\big(f_i:M_i\to N\big)_{i=1,\dots,n}$ of mutually
causally disjoint $\Loc_m$-morphisms, i.e.\ $f_i\perp f_j$ for all $i\neq j$. 
The latter assignment is required to be compatible with compositions
of tuples, unital with respect to the identities $\id:M\to M$, and equivariant with respect
to permutation actions. The AQFT is said to satisfy the time-slice axiom if it assigns to each Cauchy morphism
an $\Alg_{\mathsf{uAs}}(\TT)$-isomorphism, i.e.\
\begin{flalign}\label{eqn:AQFTpictures3}
\parbox{8cm}{
\begin{tikzpicture}[scale=0.5]
\filldraw[fill=gray!20, draw=black] (0,-8) -- (1.8,-6.2) -- (3.6,-8) -- (1.8,-9.8) -- (0,-8);
\draw (3.1,-6.7) node {\footnotesize{$N$}};
\draw[fill=gray!70, draw=black] (0,-8) -- (1.8,-7.2) -- (3.6,-8) -- (1.8,-8.8) -- (0,-8);
\draw (1.8,-6.9) node {\footnotesize{$M$}};
\draw[very thick, dashed] (0,-8)  .. controls (1,-7.8) and (2.6,-8.2) ..   (3.6,-8) node[below,midway] {\footnotesize{$\Sigma$}};
\draw[very thick, |->] (5,-8) -- (8,-8) node[midway,above] {{\footnotesize{\text{AQFT}}}};
\draw (13,-8) node {$\mathfrak{A}(M) ~\stackrel{\cong}{\longrightarrow} ~\mathfrak{A}(N)$};
\end{tikzpicture}}\quad.
\end{flalign}
\end{subequations}
We would like to stress that this definition of AQFTs comes with two different kinds
of multiplications: (1)~On every spacetime $M\in \Loc_m$ we have an associative
and unital multiplication on $\mathfrak{A}(M) \in \Alg_{\mathsf{uAs}}(\TT)$, which we denote
by $a\cdot a^\prime \in \mathfrak{A}(M)$, for all $a,a^\prime\in \mathfrak{A}(M)$. This multiplication
has its origins in quantum theory and it is physically interpreted as the (associative but noncommutative) 
product of quantum observables. (2)~For every pair of causally disjoint spacetime embeddings
$f_1:M_1\to N \leftarrow M_2:f_2$ we have another 
multiplication $\AAA(M_1)\otimes \AAA(M_2)\to \AAA(N)\,,~
a_1\otimes a_2\mapsto a_1\bullet a_2$ that takes quantum observables from causally disjoint 
$M_1$ and $M_2$ and produces a quantum observable in $\AAA(N)$. The origin of this multiplication
lies in Lorentzian geometry (not in quantum physics!) and it should be interpreted
as a kind of ``composition of causally disjoint subsystems''. The fact that $\bullet$ is by 
hypothesis an algebra morphism with respect to $\cdot$ then implies Einstein causality via the following
Eckmann-Hilton-type argument: For all $a_1\in \AAA(M_1)$ and $a_2\in \AAA(M_2)$, we compute
\begin{flalign}
\nn \AAA(f_1)(a_1)\cdot \AAA(f_2)(a_2) &= (a_1\bullet \oone_2)\cdot (\oone_1\bullet a_2)
= (a_1\cdot \oone_1)\bullet (\oone_2\cdot a_2) = (\oone_1\cdot a_1)\bullet (a_2\cdot \oone_2) \\
&= (\oone_1\bullet a_2)\cdot (a_1\bullet \oone_2) = 
\AAA(f_2)(a_2)\cdot \AAA(f_1)(a_1)\quad,\label{eqn:EckmannHilton}
\end{flalign}
i.e.\ spacelike separated quantum observables commute.
\sk

From this description it becomes rather obvious that there exists an operad 
whose algebras in the symmetric monoidal category $\TT$ are precisely AQFTs.
This operad takes the form
\begin{flalign}\label{eqn:AQFToperadspecial}
\O_{(\Loc_m,\perp)}[W^{-1}] \,:=\, \big(\P_{(\Loc_m,\perp)}\otimes_{\mathrm{BV}}^{} \mathsf{uAs}\big)[W^{-1}]\quad.
\end{flalign}
The first factor $\P_{(\Loc_m,\perp)}$ denotes the operad describing ``compositions of causally disjoint subsystems''
\eqref{eqn:AQFTpictures2}. The second factor $\mathsf{uAs}$ is the unital associative operad from \eqref{eqn:uAs}
and it describes the spacetime-wise multiplication of quantum observables. 
The symbol $\otimes_{\mathrm{BV}}^{}$ is the Boardman-Vogt tensor product of operads, which takes
care of the compatibility conditions between the two kinds of operations. The symbol 
$[W^{-1}]$ stands for a \textit{localization of operads} at the class of Cauchy morphisms $W$,
see e.g.\ \cite[Definition 2.12]{strictification} for a precise definition, which implements the 
time-slice axiom \eqref{eqn:AQFTpictures3} by universally inverting all Cauchy morphisms in
$\O_{(\Loc_m,\perp)}$. There exist more concrete alternative 
descriptions of the AQFT operad $\O_{(\Loc_m,\perp)}$ (before localization), see e.g.\
\cite[Proposition 3.9]{operadAQFT} for a closed expression
or \cite[Section 3.3]{operadAQFT} for a presentation in terms of generators and relations.
\sk

The AQFT  operad \eqref{eqn:AQFToperadspecial} generalizes in a straightforward way
to the case where $(\Loc_m,\perp)$ is replaced by an \textit{orthogonal category}
$(\CC,\perp)$, see e.g.\ \cite[Definition 3.4]{operadAQFT}, 
and $W\subseteq \mathrm{Mor}\,\CC$ is a class of morphisms in $\CC$. 
We denote the category of algebras over the operad $\O_{(\CC,\perp)}$ by
\begin{subequations}\label{eqn:AQFTcats}
\begin{flalign}
\AQFT(\CC,\perp)\,:=\, \Alg_{\O_{(\CC,\perp)}}(\TT)
\end{flalign}
and the category of algebras over the localized operad $\O_{(\CC,\perp)}[W^{-1}]$
by
\begin{flalign}
\AQFT(\CC,\perp)^{W}\,:=\, \Alg_{\O_{(\CC,\perp)}[W^{-1}]}(\TT)\quad.
\end{flalign}
\end{subequations}
Note that $\AQFT(\CC,\perp)^{\emptyset} = \AQFT(\CC,\perp)$, so the first definition
is a special case of the second one. More surprisingly, 
we shall see in Proposition \ref{prop:timeslicelocalization} below that the second definition 
is covered by the first one if one replaces $(\CC,\perp)$ by the localized
orthogonal category $(\CC[W^{-1}],\perp^W)$.
\sk

Arranging orthogonal categories,
orthogonal functors and their natural transformation into a $2$-category
$\Cat^{\perp}$, see \cite[Definition 2.1]{strictification}, allows us to upgrade
the assignment $(\CC,{\perp})\mapsto \O_{(\CC,\perp)}$ to a $2$-functor
$\O : \Cat^\perp\to \Op$ to the $2$-category of operads, morphisms and transformations, see 
\cite[Proposition 2.6]{strictification}. From this, one obtains that the assignment
$(\CC,\perp)\mapsto \AQFT(\CC,\perp)$ of AQFT categories \eqref{eqn:AQFTcats} 
is contravariantly $2$-functorial. In particular, each orthogonal functor
$F : (\CC,\perp_\CC)\to (\DD,\perp_\DD)$, i.e.\ $F(f_1)\perp_\DD F(f_2)$ for all
$f_1\perp_\CC f_2$, defines a pullback functor
\begin{flalign}\label{eqn:AQFTpullback}
F^\ast\,:\, \AQFT(\DD,\perp_\DD)\,\longrightarrow\,\AQFT(\CC,\perp_\CC)
\end{flalign}
at the level of AQFT categories. We shall illustrate below that this is useful
for passing between and comparing different flavors of AQFTs.

\paragraph{Operadic Kan extensions:} Suppose that
the symmetric monoidal category $\TT$ is closed (i.e.\ internal homs exist)
and bicomplete (i.e.\ all small limits and colimits exist).
One then might ask if the functor \eqref{eqn:AQFTpullback} admits a left and/or a right adjoint,
which in the context of ordinary category theory are known as left/right Kan extensions.
In contrast to categories, which we think of as describing $1$-to-$1$ operations, 
operads have a built-in asymmetry as they describe $n$-to-$1$ operations
but no $1$-to-$n$ operations for $n\geq 2$. This favors the existence of 
a left adjoint over a right adjoint to the functor \eqref{eqn:AQFTpullback}.
The following well-known existence result for operadic left Kan extensions
(see e.g.\ \cite[Theorem 2.11]{operadAQFT} for a spelled out proof)
has, as we shall see below, interesting consequences for AQFT.
\begin{propo}\label{prop:operadicLKE}
Suppose that $\TT$ is a bicomplete closed symmetric monoidal category.
Then, for every orthogonal functor $F : (\CC,\perp_\CC)\to (\DD,\perp_\DD)$, 
the pullback functor \eqref{eqn:AQFTpullback} admits a left adjoint, i.e.\ we have an adjunction
\begin{flalign}
\xymatrix{
F_!\,:\,\AQFT(\CC,\perp_\CC)\ar@<0.75ex>[r]~&~\ar@<0.75ex>[l] \AQFT(\DD,\perp_\DD)\,:\,F^\ast
}\quad.
\end{flalign}
In stark contrast to this, $F^\ast$ does \emph{not} in general admit a right adjoint.
\end{propo}

\begin{ex}\label{ex:Fredenhagenalgebra}
The most basic and intuitive example is given by considering
the orthogonal functor $j :(\Loc_m^{\text{\large $\diamond$}},\perp)\to (\Loc_{m},\perp)$
that describes the full orthogonal subcategory embedding of topologically
trivial spacetimes (i.e.\ diffeomorphic to $\mathbb{R}^m$, hence looking like diamonds $\diamond$) 
into the category of all spacetimes.
In this case the operadic left Kan extension 
$j_! :\AQFT(\Loc_m^{\text{\large $\diamond$}},\perp)\to \AQFT(\Loc_{m},\perp)$
extends AQFTs that are defined only on diamonds to all spacetimes. This is a refinement
of Fredenhagen's universal algebra construction \cite{Fredenhagen}, see \cite[Section 5]{operadAQFT}
for a detailed comparison.
\end{ex}

\begin{rem}
Even though operadic \textit{right} Kan extension often do not exist,
there are some important and interesting exceptions. For instance, orbifoldization
of AQFTs \cite[Section 4.5]{operadAQFT} and the 
extraction of the chiral observables of a $2$-dimensional conformal AQFT
\cite[Section 5]{CLoc2} are formalized by right adjoints. 
\end{rem}

\paragraph{Classification results in low dimensions:}
Operadic techniques are also vital to obtain a deeper
understanding of the time-slice axiom. Let $(\CC,\perp)$ be an 
orthogonal category and $L : \CC\to \CC[W^{-1}]$ a localization
functor for the underlying category $\CC$ at a set $W\subseteq \mathrm{Mor}\,\CC$
of morphisms, which we think of as an abstraction of Cauchy morphisms.
We upgrade $L$ to an orthogonal functor $L : (\CC,\perp)\to (\CC[W^{-1}],\perp^W)$
by endowing the localized category with the push-forward orthogonality relation $\perp^W$,
i.e.\ the minimal orthogonality relation such that $L(f_1) \perp^W L(f_2)$
for all $f_1\perp f_2$. Using ($2$-)functoriality of the assignment of AQFT operads,
we obtain an operad morphism $\O_L : \O_{(\CC,\perp)}\to \O_{(\CC[W^{-1}],\perp^W)}$
which by \cite[Proposition 2.14]{strictification} provides a model
for the localized operad $\O_{(\CC,\perp)}[W^{-1}]$. In simpler words,
localizing the AQFT operad is equivalent to localizing the underlying orthogonal category.
A direct consequence is the following
\begin{propo}\label{prop:timeslicelocalization}
For any orthogonal category $(\CC,\perp)$ and subset $W\subseteq\mathrm{Mor}\,\CC$, we have an equivalence
of categories
\begin{flalign}
\AQFT(\CC,\perp)^W \,\simeq \, \AQFT\big(\CC[W^{-1}],\perp^W\big)\quad.
\end{flalign}
\end{propo}

This proposition paves the way to classification results for AQFTs in low dimensions.
\begin{ex}\label{ex:Loc1}
This example is based on \cite[Appendix A]{strictification}.
The localization of the orthogonal category $(\Loc_1,\emptyset)$ of 
$1$-dimensional connected spacetimes at all Cauchy morphisms is given by
\begin{flalign}
\big(\Loc_1[W^{-1}],\emptyset\big)\,\simeq \,\big(\mathbf{B}\bbR,\emptyset\big)\quad,
\end{flalign}
where $\mathbf{B}\bbR$ denotes the groupoid with a single object, denoted by
$\parbox{0.3em}{\begin{tikzpicture}[scale=0.18]
\draw[ thick] (0,0) -- (0,1.5);
\end{tikzpicture}}$, and morphisms given by the additive group $(\bbR,+,0)$. Using Proposition 
\ref{prop:timeslicelocalization}, one then finds that an AQFT on $(\Loc_1,\emptyset)$
that satisfies the time-slice axiom is equivalent to the datum 
\begin{flalign}
\xymatrix{
\AAA(\,\parbox{0.3em}{\begin{tikzpicture}[scale=0.18]
\draw[ thick] (0,0) -- (0,1.5);
\end{tikzpicture}})\ar@(ul,dl)_-{\bbR}
}~\in~\mathbf{Fun}\big(\mathbf{B}\bbR,\Alg_{\mathsf{uAs}}(\TT)\big)
\end{flalign}
of a single algebra endowed with an $\bbR$-action, which we interpret as time-evolution. Hence,
$1$-dimensional AQFTs describe precisely the same content as ordinary quantum mechanics.
\end{ex}
\begin{ex}\label{ex:CLoc2}
The previous example can be generalized to $2$-dimensional \textit{conformal} AQFTs that
are defined on the orthogonal category $(\mathbf{CLoc}_2,\perp)$, see \cite[Section 3]{CLoc2}.
The result is that an AQFT on $(\mathbf{CLoc}_2,\perp)$
that satisfies the time-slice axiom is equivalent to the datum
\begin{flalign}
\xymatrix@C=6em{
\ar@(ul,dl)_-{\mathrm{Emb}(\bbR)^2} \AAA(\parbox{0.75em}{\begin{tikzpicture}[scale=0.12]
\draw (0,0) -- (1,1) -- (2,0) -- (1,-1) -- (0,0);
\end{tikzpicture}}) 
 \ar[r]^-{\mathrm{Emb}(\parbox{0.7em}{\begin{tikzpicture}[scale=0.12]
\draw (0,0) -- (1,1) -- (2,0) -- (1,-1) -- (0,0);
\end{tikzpicture}}\,,\,\parbox{0.7em}{\begin{tikzpicture}[scale=0.065]
\draw (0,0) ellipse (1.25 and 0.5);
\draw (-1.25,0) -- (-1.25,-3.5);
\draw (-1.25,-3.5) arc (180:360:1.25 and 0.5);
\draw [densely dotted] (-1.25,-3.5) arc (180:360:1.25 and -0.5);
\draw (1.25,-3.5) -- (1.25,0);  
\end{tikzpicture}}
)}~&~\AAA(\parbox{0.65em}{\begin{tikzpicture}[scale=0.065]
\draw (0,0) ellipse (1.25 and 0.5);
\draw (-1.25,0) -- (-1.25,-3.5);
\draw (-1.25,-3.5) arc (180:360:1.25 and 0.5);
\draw [densely dotted] (-1.25,-3.5) arc (180:360:1.25 and -0.5);
\draw (1.25,-3.5) -- (1.25,0);  
\end{tikzpicture}}) \ar@(ur,dr)^{\mathrm{Diff}(\mathbb{S}^1)^2}
}
\end{flalign}
of two algebras, one for the Minkowski spacetime 
$\AAA(\parbox{0.75em}{\begin{tikzpicture}[scale=0.12]
\draw (0,0) -- (1,1) -- (2,0) -- (1,-1) -- (0,0);
\end{tikzpicture}})$ and one for the flat cylinder 
$\AAA(\parbox{0.65em}{\begin{tikzpicture}[scale=0.065]
\draw (0,0) ellipse (1.25 and 0.5);
\draw (-1.25,0) -- (-1.25,-3.5);
\draw (-1.25,-3.5) arc (180:360:1.25 and 0.5);
\draw [densely dotted] (-1.25,-3.5) arc (180:360:1.25 and -0.5);
\draw (1.25,-3.5) -- (1.25,0);  
\end{tikzpicture}})$, together with the displayed actions of the embedding monoid $\mathrm{Emb}(\bbR)^2$,
of the diffeomorphism group $\mathrm{Diff}(\mathbb{S}^1)^2$ and of the conformal embeddings
$\mathrm{Emb}(\parbox{0.7em}{\begin{tikzpicture}[scale=0.12]
\draw (0,0) -- (1,1) -- (2,0) -- (1,-1) -- (0,0);
\end{tikzpicture}}\,,\,\parbox{0.7em}{\begin{tikzpicture}[scale=0.065]
\draw (0,0) ellipse (1.25 and 0.5);
\draw (-1.25,0) -- (-1.25,-3.5);
\draw (-1.25,-3.5) arc (180:360:1.25 and 0.5);
\draw [densely dotted] (-1.25,-3.5) arc (180:360:1.25 and -0.5);
\draw (1.25,-3.5) -- (1.25,0);  
\end{tikzpicture}}
)$. (These actions have to be compatible with causal disjointness $\perp$.)
This relates $2$-dimensional conformal AQFT
and representation theoretic approaches to conformal QFTs.
\end{ex}

\paragraph{Comparison with factorization algebras:} Factorization algebras
\cite{CG1,CG2} provide a different axiomatization of QFT that is applicable
to various geometric contexts, such as topological, holomorphic, Riemannian, 
and also Lorentzian geometry. In the Lorentzian context, the general definition 
of Costello and Gwilliam can be refined slightly to the more appropriate 
variant of \textit{time-orderable prefactorization algebras} \cite{FAvsAQFT},
which emphasize time-orderability of spacetime regions instead of the coarser 
concept of disjointness. The relevant geometric definitions are as follows:
\begin{itemize}
\item[(iii)] A tuple $\big(f_i : M_i\to N\big)_{i=1,\dots,n}$ of $\Loc_m$-morphisms
is called \textit{time-ordered} if $J_N^+(f_i(M_i))\cap f_j(M_j) = \emptyset$, for all $i<j$.

\item[(iv)] A tuple is called \textit{time-orderable} if there exists a permutation $\rho\in\Sigma_n$ 
(called time-ordering permutation) such that the permuted tuple 
$\big(f_{\rho(i)} : M_{\rho(i)}\to N\big)_{i=1,\dots,n}$ is time-ordered.
\end{itemize}

A time-orderable prefactorization algebra (tPFA) on $\Loc_m$  is given by the assignment
\begin{subequations}\label{eqn:tPFApictures}
\begin{flalign}\label{eqn:tPFApictures1}
\parbox{8cm}{
\begin{tikzpicture}[scale=0.5]
\filldraw[fill=gray!20, draw=black] (0,0) -- (1.8,1.8) -- (3.6,0) -- (1.8,-1.8) -- (0,0);
\draw (3.1,1.3) node {\footnotesize{$M$}};
\draw[very thick, |->] (5,0) -- (8,0) node[midway,above] {{\footnotesize{\text{tPFA}}}};
\draw (11,0) node {$\mathfrak{F}(M) \in \TT$};
\end{tikzpicture}}
\end{flalign}
of an object in $\TT$ to each spacetime $M\in \Loc_m$, together with the assignment
\begin{flalign}\label{eqn:tPFApictures2}
\parbox{8cm}{
\begin{tikzpicture}[scale=0.5]
\filldraw[fill=gray!20, draw=black] (0,-4) -- (1.8,-2.2) -- (3.6,-4) -- (1.8,-5.8) -- (0,-4);
\draw (3.1,-2.7) node {\footnotesize{$N$}};
\draw[fill=gray!70, draw=black] (0.2,-4) -- (0.7,-3.5) -- (1.2,-4) -- (0.7,-4.5) -- (0.2,-4);
\draw (0.7,-4) node {\tiny{$M_1$}};
\draw[fill=gray!70, draw=black] (2.4,-4) -- (2.9,-3.5) -- (3.4,-4) -- (2.9,-4.5) -- (2.4,-4);
\draw (2.9,-4) node {\tiny{$M_2$}};
\draw[fill=gray!70, draw=black] (1.3,-3) -- (1.8,-2.5) -- (2.3,-3) -- (1.8,-3.5) -- (1.3,-3);
\draw (1.8,-3) node {\tiny{$M_n$}};
\draw (1.8,-4.5) node {{$\vdots$}};
\draw[very thick, |->] (5,-4) -- (8,-4) node[midway,above] {{\footnotesize{\text{tPFA}}}};
\draw (13,-4) node {$\bigotimes\limits_{i=1}^n \FFF(M_i) ~\longrightarrow ~\FFF(N)$};
\end{tikzpicture}}
\end{flalign}
\end{subequations}
of a $\TT$-morphism to each time-orderable tuple $\big(f_i : M_i\to N\big)_{i=1,\dots,n}$. 
The latter assignment is subject to 
compatibility with compositions of tuples, unitality and equivariance with respect to permutation actions.
A tPFA is said to satisfy the time-slice axiom if, in analogy to \eqref{eqn:AQFTpictures3},
it assigns to each Cauchy morphism a $\TT$-isomorphism $\FFF(M)\stackrel{\cong}{\longrightarrow} \FFF(N)$.
\sk

Comparing this definition with the one of AQFTs in \eqref{eqn:AQFTpictures},
one realizes that tPFAs have less spacetime-wise algebraic structure (one 
assigns plain objects in $\TT$ and not unital associative algebras) but more 
composition products since there are more time-orderable tuples than mutually causally disjoint ones.
In the context of free QFTs, it was observed first in \cite{GR1} that, by means of the time-slice axiom,
one can construct suitable spacetime-wise multiplications in a tPFA. This observation
was then generalized in \cite{FAvsAQFT} to a model-independent statement.
\sk

In order to state these results, let us denote by $\mathsf{t}\mathcal{P}_m[W^{-1}]$ the operad
encoding time-orderable prefactorization algebras that satisfy the time-slice axiom
and by
\begin{flalign}
\tPFA(\Loc_m)^W\,:=\, \Alg_{\mathsf{t}\mathcal{P}_m[W^{-1}]}^{}(\TT)
\end{flalign}
its category of algebras. It is shown in \cite[Remark 5.2]{FAvsAQFT} that there exists
an operad morphism $\Phi : \mathsf{t}\mathcal{P}_m[W^{-1}]\to \O_{(\Loc_m,\perp)}[W^{-1}]$ to the AQFT operad,
which defines a pullback functor
\begin{flalign}\label{eqn:AQFTtotPFA}
\Phi^{\ast} \,:\, \AQFT(\Loc_m,\perp)^W~\longrightarrow~\tPFA(\Loc_m)^W\quad,
\end{flalign}
i.e.\ every AQFT has an underlying tPFA. Restricting to the full subcategories
of \textit{additive objects} \cite[Definitions 2.11 and 2.16]{FAvsAQFT}, i.e.\ 
tPFAs/AQFTs whose value on any $M\in \Loc_m$ agrees with the 
colimit of the values on the category of all relatively compact $U\to M$ in $\Loc_m$, one
obtains the following main result.
\begin{theo}\label{theo:comparison}
There exists an isomorphism of categories
\begin{flalign}
\xymatrix{
\Phi_!\,:\,\tPFA(\Loc_m)^{W,\,\mathrm{add}}\ar@<1.1ex>[r]_-{\cong}~&~\ar@<1.1ex>[l] \AQFT(\Loc_m,\perp)^{W,\,\mathrm{add}}\,:\,\Phi^\ast
}\quad.
\end{flalign}
The functor $\Phi_!$ admits an explicit description that is spelled out in
\cite[Section 3]{FAvsAQFT}.
\end{theo}

%%%%%%%%%%%%%%%%%%%%%%%%%%%%%%%%%%%%%%%%%%%%%%%%
%%%%%%%%%%%%%%%%%%%%%%%%%%%%%%%%%%%%%%%%%%%%%%%%

\section{\label{sec:homotopy}Homotopy theory}

\paragraph{Motivation from BRST/BV:} The construction of gauge-theoretic examples
of AQFTs is usually carried out in the context of the BRST/BV formalism, see e.g.\ \cite{FR1,FR2}.
This leads to AQFTs that take values in the symmetric monoidal category 
$\TT=\Ch_\bbK$ of cochain complexes of vector spaces over a field $\bbK$ of characteristic $0$ 
(typically $\bbK=\bbC$), which in short we call \textit{dg-AQFTs}. 
Working with cochain complexes provides a natural habitat
for the ghost fields and antifields, both of which have non-zero cohomological degree, 
as well as the BRST and BV differentials. Looking closer into these constructions, 
one observes that (the isomorphism type of) their output depends on the auxiliary
choices that one makes along the way, such as gauge fixings and auxiliary fields. 
This unpleasant and seemingly unphysical feature is overcome by realizing that 
isomorphisms are not the right concept to compare two cochain complexes,
but rather one should use the weaker concept of \textit{quasi-isomorphisms}. So, loosely speaking, 
the best one can hope for is that the BRST/BV formalism constructs from a given physical input (say, a classical
action functional) a dg-AQFT $\AAA$ that is determined uniquely up to quasi-isomorphisms. In particular, making
different choices for, e.g., gauge fixings and/or auxiliary fields should lead
to quasi-isomorphic outputs $\AAA^\prime\stackrel{\sim}{\longrightarrow}\AAA$.
\sk

Ensuring that the output of a construction does not depend (up to quasi-isomorphisms) on 
auxiliary intermediate choices is hard to do by hand, hence one needs mathematical technology
that takes care of these issues. Model category theory \cite{Hovey,Hirschhorn}, which will be briefly discussed
in the next paragraph, provides powerful tools for this purpose. ($\infty$-category theory
provides an alternative and more modern approach. We prefer to work with model categories
since they are more explicit and hence better suited for concrete computations.)

\paragraph{The basic idea behind model categories:} A \textit{model category} is a bicomplete
category $\MM$ together with the choice of three classes of morphisms, called \textit{weak equivalences},
\textit{fibrations} and \textit{cofibrations}, that satisfy various axioms, see e.g.\ \cite[Section 1.1]{Hovey}.
The weak equivalences should be thought of as a generalization of the concept of isomorphisms 
in ordinary category theory. In particular, two weakly equivalent objects in a model category are 
considered to be ``the same''. The role of the fibrations and cofibrations is less direct, but 
still important, as they are used to prove existence of, and
determine models for, \textit{derived functors}.
\sk

To motivate the latter concept, let us start with the important (but trivial) observation
that a functor $F : \MM\to\NN$ between two model categories in general \textit{does not} 
preserve weak equivalences, which is manifestly inconsistent with the basic idea that weakly equivalent
objects should be regarded as being ``the same''. For certain types of functors, 
this issue can be resolved by deforming them to derived functors. 
An important case where this is possible is when the functor is part of a \textit{Quillen adjunction}, 
which is an adjunction
\begin{flalign}
\xymatrix{
F\,:\,\MM\ar@<0.75ex>[r]~&~\ar@<0.75ex>[l] \NN \,:\,G
}
\end{flalign}
between two model categories $\MM$ and $\NN$, such that the left
adjoint $F$ preserves cofibrations and the right adjoint $G$ preserves
fibrations. (See also \cite[Proposition 8.5.3]{Hirschhorn} 
for a list of equivalent characterizations that is useful in practice.) Choosing a cofibrant
resolution $(Q : \MM\to \MM,\, q:Q\Rightarrow \id_\MM)$ for $\MM$
and a fibrant resolution $(R:\NN\to \NN, \,r: \id_{\NN}\Rightarrow R)$ for $\NN$,
which exist by the model category axioms, one defines the derived functors
\begin{flalign}
\mathbb{L} F\,:=\,F\circ Q\,:\,\MM\,\longrightarrow\,\MM\quad,\qquad
\bbR G\,:=\,G\circ R\,:\,\NN\,\longrightarrow\,\NN\quad.
\end{flalign}
The crucial feature of derived functors is that they preserve weak equivalences,
i.e.\ they are compatible with the main philosophy that weakly equivalent
objects are ``the same''.

\paragraph{dg-AQFT model categories and Quillen adjunctions:} Motivated
by the BRST/BV formalism, let us consider dg-AQFTs, i.e.\ AQFTs that take values in the bicomplete closed
symmetric monoidal category $\TT=\Ch_\bbK$ of cochain complexes of vector spaces
over a field $\bbK$ of characteristic $0$. The latter is a \textit{symmetric monoidal model
category}, with weak equivalences given by the quasi-isomorphisms, fibrations given by degree-wise
surjective cochain maps and cofibrations determined implicitly by a lifting property, see e.g.\
\cite[Sections 2.3 and 4.2]{Hovey} for the details. Using fundamental
insights by Hinich \cite{Hinich}, we obtain the following important structural 
result for dg-AQFTs, which was first reported in \cite{hAQFT}.
\begin{theo}\label{theo:AFTmodelstructure}
\begin{itemize}
\item[(1)] For any orthogonal category $(\CC,\perp)$, the category
\begin{flalign}
\dgAQFT(\CC,\perp)\,:=\,\Alg_{\O_{(\CC,\perp)}}^{}\big(\Ch_\bbK\big)
\end{flalign}
of cochain complex valued AQFTs is a model category for the following projective model structure: 
A morphism $\zeta : \AAA\to\BBB$ is a weak equivalence (resp.\ fibration)
if each component $\zeta_M : \AAA(M)\to \BBB(M)$, for $M\in \CC$, 
is a quasi-isomorphism (resp.\ degree-wise surjective cochain map). The cofibrations are determined
by the left lifting property with respect to all acyclic fibrations, i.e.\ morphisms that are simultaneously 
weak equivalences and fibrations.

\item[(2)] For any orthogonal functor $F : (\CC,\perp_\CC)\to (\DD,\perp_\DD)$, the adjunction
\begin{flalign}
\xymatrix{
F_!\,:\,\dgAQFT(\CC,\perp_\CC)\ar@<0.75ex>[r]~&~\ar@<0.75ex>[l] \dgAQFT(\DD,\perp_\DD)\,:\,F^\ast
}
\end{flalign}
from Proposition \ref{prop:operadicLKE} is a Quillen adjunction.
\end{itemize}
\end{theo}

The model categories from item (1) provide a natural context
for receiving AQFT model constructions via the BRST/BV formalism \cite{FR1,FR2}
and, very importantly, for ensuring that the outputs do not depend on
intermediate choices such as gauge fixings and/or auxiliary fields.
The latter aspects have been studied in detail in the context
of linear quantum gauge theories, see in particular \cite{BSlinear,linYM,FreeFAvsAQFT}.
The Quillen adjunctions from item (2) lead to interesting applications, such as
derived variants of the local-to-global extensions discussed in Example \ref{ex:Fredenhagenalgebra}.
Toy-examples for the latter, which illustrate the rich interplay between
gauge-theoretic structures and derived functors, are described in \cite[Appendix A]{hAQFT}.

\paragraph{Homotopy-coherent AQFTs:} The model category $\dgAQFT(\CC,\perp)$ 
from Theorem \ref{theo:AFTmodelstructure} describes dg-AQFTs 
in which all algebraic structures compose strictly,
i.e.\ without non-trivial homotopy coherence data. One may therefore ask
whether there exists a potentially more general and flexible concept
of \textit{homotopy-coherent AQFTs} which relates to dg-AQFT
such as, e.g., $A_\infty$-algebras relate to associative dg-algebras
and $L_\infty$-algebras relate to dg-Lie algebras. Operad theory
provides the necessary tools to investigate and answer this question, see e.g.\ 
\cite{BergerMoerdijk,Hinich} for more details. In the present context, a homotopy-coherent
AQFT is then, by definition, a homotopy algebra over the
dg-operad $\O_{(\CC,\perp)}\otimes\bbK\in\mathbf{dgOp}_\bbK$ that is obtained
by turning all sets of operations into cochain complexes via 
$(-)\otimes \bbK: \Set\to\Ch_\bbK\,,~S\mapsto S\otimes \bbK := \bigoplus_{s\in S}\bbK$.
Again by definition, a homotopy algebra over $\O_{(\CC,\perp)}\otimes\bbK$ is
an ordinary algebra over {\it any} choice of cofibrant (or $\Sigma$-cofibrant, to obtain smaller models)
resolution $\O_{(\CC,\perp)\,\infty} \stackrel{\sim}{\longrightarrow} \O_{(\CC,\perp)}\otimes\bbK$
of the given dg-operad. This means that homotopy-coherent 
AQFTs are described by the category
\begin{flalign}
\dgAQFT_{\infty}(\CC,\perp)\,:=\,\Alg_{\O_{(\CC,\perp)\,\infty}}^{}\big(\Ch_\bbK\big)\quad,
\end{flalign}
which we again endow with the projective model structure.
\sk

It turns out that homotopy-coherent AQFTs \textit{do not}
carry additional content when working, as we always do, over a field $\bbK$ of characteristic $0$.
This is a consequence of general rectification theorems, see e.g.\ \cite[Theorem 4.1]{BergerMoerdijk}
and \cite[Theorem 2.4.5]{Hinich}.
\begin{theo}
Let $(\CC,\perp)$ be any orthogonal category and 
$q : \O_{(\CC,\perp)\,\infty} \stackrel{\sim}{\longrightarrow} \O_{(\CC,\perp)}\otimes\bbK$
any ($\Sigma$-)cofibrant resolution of the AQFT dg-operad. Then the induced Quillen adjunction
\begin{flalign}
\xymatrix{
q_!\,:\,\dgAQFT_\infty(\CC,\perp)\ar@<0.75ex>[r]_-{\sim}~&~\ar@<0.75ex>[l] \dgAQFT(\CC,\perp)\,:\,q^\ast
}
\end{flalign}
is a Quillen equivalence. In particular, every homotopy-coherent
AQFT $\AAA_\infty\in \dgAQFT_\infty(\CC,\perp)$ admits a strictification
to a dg-AQFT that is given by the derived unit $\AAA_\infty \stackrel{\sim}{\rightsquigarrow} \AAA_{\mathrm{st}}
:= \bbR q^\ast \,\mathbb{L} q_!(\AAA_\infty)$.
\end{theo}

\begin{open}
Recalling from \eqref{eqn:AQFToperadspecial} the presentation of
the AQFT operad as a Boardman-Vogt tensor product 
$\O_{(\CC,\perp)} = \P_{(\CC,\perp)}\otimes_{\mathrm{BV}}^{}\mathsf{uAs}$, there is
a pathway to introduce a potentially richer concept
of homotopy-coherent AQFTs. Regarding both factors, i.e.\ $\P_{(\CC,\perp)}$ and 
$\mathsf{uAs}$, as $\infty$-operads in the sense of \cite{LurieHA},
we can form their $\infty$-operadic tensor product $\P_{(\CC,\perp)}\otimes_{\infty}^{} \mathsf{uAs}$.
This $\infty$-operad is a priori different from the operads obtained by cofibrant resolutions.
Indeed, from Lurie-Dunn additivity \cite[Theorem 5.1.2.2]{LurieHA}, one finds that 
$\mathsf{uAs}\otimes_{\infty}^{}\mathsf{uAs}\simeq \mathbb{E}_2$ is the homotopically
non-trivial $\mathbb{E}_2$-operad, while $(\mathsf{uAs}\otimes_{\mathrm{BV}}^{}\mathsf{uAs})_\infty \simeq
\mathbb{E}_\infty$ is the homotopically trivial $\mathbb{E}_\infty$-operad.
In the AQFT context, where Einstein causality is implemented by 
an Eckmann-Hilton-type argument \eqref{eqn:EckmannHilton}, we believe that the 
$\infty$-operad $\P_{(\CC,\perp)}\otimes_{\infty}^{} \mathsf{uAs}$ describes
an $\mathbb{E}_2$-variant of the causality axiom. A detailed study of this $\infty$-operad
is an interesting open problem.
\end{open}

\paragraph{Strictification of the time-slice axiom:} In the previous paragraph
we have silently avoided talking about the time-slice axiom.
Recall that this is implemented in the $1$-categorical setting by
a localization of operads, see \eqref{eqn:AQFToperadspecial}. In 
the model categorical context there exist two a priori different 
options for the time-slice axiom.
\begin{defi}
A dg-AQFT $\AAA\in\dgAQFT(\CC,\perp)$
is said to satisfy the \textit{strict time-slice axiom} for $W\subseteq \mathrm{Mor}\,\CC$
if it sends each $W$-morphism $f: M\to N$ to an \textit{isomorphism}
$\AAA(f): \AAA(M)\stackrel{\cong}{\longrightarrow} \AAA(N)$.
It is said to satisfy the \textit{homotopy time-slice axiom}
if it sends each $W$-morphism $f: M\to N$ to a \textit{quasi-isomorphism}
$\AAA(f): \AAA(M)\stackrel{\sim}{\longrightarrow} \AAA(N)$. 
\end{defi}
Clearly, the strict
time-slice axiom is a special case of the homotopy variant. It is 
important to emphasize that the typical examples of dg-AQFTs
arising from the BRST/BV formalism \cite{FR1,FR2,linYM,FreeFAvsAQFT}
satisfy the homotopy time-slice axiom, but \textit{not} 
the strict one. Conceptually, this is not at all a problem, since
quasi-isomorphic and not only isomorphic objects should be regarded as being ``the same''.
However, from a more practical point of view, there are complications. For instance, 
certain constructions in AQFT, such as determining the relative Cauchy evolution (RCE) 
and the associated stress-energy tensor \cite{BFV}, become much more involved for theories
that satisfy only the homotopy time-slice axiom, see e.g.\ \cite{BFSRCE}.
This motivates and justifies the search for strictification theorems for the homotopy time-slice axiom.
In order to set the stage, we first have to introduce
model categories for both variants of the time-slice axiom.
In the strict case, we use Proposition \ref{prop:timeslicelocalization} 
and endow
\begin{flalign}
\dgAQFT(\CC,\perp)^W\,:=\,\Alg_{\O_{(\CC[W^{-1}],\perp^W)}}^{}\big(\Ch_\bbK\big)
\end{flalign}
with the projective model structure. In the homotopy case, we follow
\cite[Section 3]{Carmona} and consider
\begin{flalign}
\dgAQFT(\CC,\perp)^{\mathrm{ho}W}\,:=\, \Alg_{\O_{(\CC,\perp)}[W^{-1}]^\infty}^{}\big(\Ch_\bbK\big)\,\simeq\,
\mathcal{L}_{\widehat{W}}\dgAQFT(\CC,\perp)\quad.
\end{flalign}
The first step involves the \textit{homotopical localization}
$\O_{(\CC,\perp)}[W^{-1}]^\infty\in\mathbf{dgOp}_\bbK$ of the AQFT dg-operad
at $W$, whose category of algebras is endowed with the projective model structure.
The second step uses the Quillen equivalence from \cite[Theorem 3.13]{Carmona}
to identify the left-hand side with the left Bousfield localization of
the projective model category from Theorem \ref{theo:AFTmodelstructure} 
at a suitable class of morphisms $\widehat{W}$ that is determined from $W$.
Using the localization operad morphism $\O_L : \O_{(\CC,\perp)}\to \O_{(\CC[W^{-1}],\perp^W)}$,
we obtain by \cite[Proposition 2.24]{strictification} a Quillen adjunction
\begin{flalign}\label{eqn:timesliceadjunction}
\xymatrix{
L_!\,:\,\dgAQFT(\CC,\perp)^{\mathrm{ho}W}\ar@<0.75ex>[r]~&~\ar@<0.75ex>[l] \dgAQFT(\CC,\perp)^W\,:\,L^\ast
}
\end{flalign}
that allows us to compare the strict and homotopy time-slice axioms.
The following result is proven in \cite[Theorem 3.6]{strictification}.
\begin{theo}\label{theo:localization}
Suppose that $L : (\CC,\perp)\to (\CC[W^{-1}],\perp^W)$ is a reflective orthogonal localization,
which means that the orthogonal localization functor $L$ admits a fully faithful right adjoint orthogonal functor
$i : (\CC[W^{-1}],\perp^W)\to (\CC,\perp)$. Then  \eqref{eqn:timesliceadjunction} is a Quillen equivalence.

In particular, in this case every dg-AQFT $\AAA\in \dgAQFT(\CC,\perp)^{\mathrm{ho}W}$ satisfying
the homotopy time-slice axiom admits a strictification to a dg-AQFT satisfying
the strict time-slice axiom that is given by the underived unit 
$\AAA\stackrel{\sim}{\to}\AAA_{\mathrm{st}}:=L^\ast\,i^\ast(\AAA)$.
\end{theo}

\begin{ex}
The following types of AQFTs are covered by this theorem, see \cite[Example 3.4]{strictification}
for more details: (i)~$1$-dimensional dg-AQFTs on $\Loc_1$, (ii)~$2$-dimensional conformal
dg-AQFTs on $\mathbf{CLoc}_2$, and (iii)~Haag-Kastler-type dg-AQFTs on the slice category $\Loc_m/M$.
In particular, (i) and (ii) imply that the classification results from Examples
\ref{ex:Loc1} and \ref{ex:CLoc2} generalize to the case of dg-AQFTs satisfying the homotopy time-slice axiom.
\end{ex}

\begin{open}
It is currently unclear to us if there exist strictification
theorems, in the form of Theorem \ref{theo:localization} or other forms, 
for the higher-dimensional case of $(\Loc_m,\perp)$, for $m\geq 2$, 
with $W$ the Cauchy morphisms. 
\end{open}

\begin{rem}
The existence of (at least some) strictification theorems 
for the homotopy time-slice axiom for dg-AQFTs 
seems to be a phenomenon that is linked to the $1$-dimensional
nature of time-evolution in Lorentzian geometry.
Analogous strictification results are clearly \textit{not} available
for topological QFTs, when formulated as locally constant factorization algebras.
Indeed, locally constant factorization algebras on $\bbR^m$ 
are equivalent to algebras over the homotopically non-trivial 
$\mathbb{E}_m$-operads \cite[Theorem 5.4.5.9]{LurieHA}, hence they can not 
be strictified for $m\geq 2$.
\end{rem}

\paragraph{Construction of examples:} The reader might now ask
how one can construct explicit examples of dg-AQFTs on $\Loc_m$. 
In this paragraph we shall sketch the construction of
linear (higher) gauge-theoretic models, which form the simplest class
of dg-AQFTs. The technical details can be found in \cite{FreeFAvsAQFT}.
\sk

This construction starts from similar input as the one of free factorization algebras
by Costello and Gwilliam \cite{CG1}, however it deviates in later steps.
As basic input, we take the following
\begin{defi}
A \textit{free BV theory on $M\in \Loc_m$} is a tuple
$\big(F_M, Q_M, (-,-)_M\big)$ consisting of a complex
of linear differential operators $(F_M,Q_M)$ and a compatible $(-1)$-shifted
fiber metric $(-,-)_M$. A \textit{free BV theory on $\Loc_m$} is a natural family
$\big(F_M, Q_M, (-,-)_M\big)_{M\in\Loc_m}$ of free BV theories on all $M\in \Loc_m$.
\end{defi}

\begin{ex}\label{ex:linYM} 
The prime example to keep in mind is linear Yang-Mills theory.
In this case the complex of linear differential operators reads as
\begin{flalign}
(F_M,Q_M)\,=\,\Big(\xymatrix{
\stackrel{(-1)}{\Lambda^0  M} \ar[r]^{\dd}~&~\stackrel{(0)}{\Lambda^1 M} \ar[r]^-{\delta\,\dd}~&~
\stackrel{(1)}{\Lambda^1 M} \ar[r]^-{\delta}~&~\stackrel{(2)}{\Lambda^0 M}
}\Big)\quad,
\end{flalign}
where $\Lambda^p M$ denotes the $p$-th exterior power of the cotangent bundle (i.e.\ $p$-forms),
$\dd$ is the de Rham differential and $\delta$ the codifferential. As typical for the BV formalism,
the underlying $\bbZ$-graded vector bundle encodes the ghost fields $c$ (in degree $-1$), 
the fields $A$ (in degree $0$), the antifields $A^\ddagger$ (in degree $1$) and 
the antifields for ghosts $c^\ddagger$ (in degree $2$).
The differential $Q_M$ encodes the action of gauge transformations via $\dd$
as well as the dynamics given by the linear Yang-Mills operator $\delta\,\dd$.
The $(-1)$-shifted fiber metric $(-,-)_M: F_M\otimes F_M\to M\times \bbR[-1]$
is given by the pairings $(A^\ddagger,A)_M = \ast^{-1}(A^\ddagger \wedge \ast A)$ and 
$(c^\ddagger,c)_M =  -\ast^{-1}(c^\ddagger \wedge \ast c)$ between fields/ghosts and their antifields, where $\ast$ denotes
the Hodge operator.
This captures the $(-1)$-shifted symplectic structure and its dual shifted Poisson structure (the antibracket) 
from the BV formalism.
More examples, including higher gauge-theoretic ones, can be found in \cite[Examples 3.6, 3.7 and 3.8]{FreeFAvsAQFT}.
\end{ex}

Associated to any free BV theory on $\Loc_m$ is a time-orderable dg-prefactorization algebra
\begin{flalign}\label{eqn:tPFAconstruction}
\xymatrix@C=3em{
\big(F_M, Q_M, (-,-)_M\big)_{M\in\Loc_m}\ar@{~>}[rr]^-{\text{BV quantization}}~&&~ \FFF\in\mathbf{dgtPFA}(\Loc_m)
}\quad,
\end{flalign}
which can be constructed via BV quantization, see \cite{CG1} and also \cite{FreeFAvsAQFT}
for the Lorentzian context. However, this setup is insufficient to construct a dg-AQFT,
which requires the additional input of retarded $G^+$ and advanced $G^-$ Green's homotopies \cite{GreenHyp}, 
generalizing the concept of retarded and advanced Green's operators. 
Informally, $G^\pm$ is a (pseudo-)natural family 
\begin{flalign}
\Big\{G^\pm_K\,\in \big[\Gamma^\infty_K(F_M), \Gamma^\infty_{J^\pm_M(K)}(F_M)\big]^{-1}\,:\, K\subseteq M\text{ compact}\, \Big\}
\end{flalign}
of homotopies with support properties precisely as retarded/advanced Green's operators
that trivialize the inclusion cochain maps
$\mathrm{incl}=\partial G_K^\pm  : \Gamma^\infty_K(F_M)\to \Gamma^\infty_{J^\pm_M(K)}(F_M)$ 
for all compacta $K \subseteq M$.
The precise definition (including pseudo-naturality) is more involved and can be found in
\cite[Definition 3.5]{GreenHyp}. In this paper many pleasant properties
of retarded/advanced Green's homotopies (generalizing the usual properties of retarded/advanced Green's operators \cite{BGP,Bar}) have been proven. Most notably, we have
\begin{theo}\label{theo:Greenexistence}
\begin{itemize}
\item[(1)] \emph{Uniqueness:} The Kan complex of retarded/advanced Green's homotopies on $(F_M,Q_M)$ 
is either empty or contractible. See \cite[Proposition 3.9]{GreenHyp}.

\item[(2)] \emph{Existence:} Suppose that $(F_M,Q_M)$ admits a \textit{Green's witness},
i.e.\ a degree $-1$ differential operator $W_M$ such that $P_M := Q_M\,W_M + W_M \,Q_M$
is a Green hyperbolic differential operator. Then $(F_M,Q_M)$ 
admits both a retarded and an advanced Green's homotopy. See \cite[Theorem 4.8]{GreenHyp}.
\end{itemize}
\end{theo}
\begin{ex}
The linear Yang-Mills complex from Example \ref{ex:linYM} admits
a natural Green's witness given by
\begin{flalign}
W_M \,=\, \Big(\xymatrix{
\stackrel{(-1)}{\Lambda^0  M} ~&~\ar[l]_-{\delta}\stackrel{(0)}{\Lambda^1 M} ~&~\ar[l]_-{\id}
\stackrel{(1)}{\Lambda^1 M} ~&~ \ar[l]_-{\dd}\stackrel{(2)}{\Lambda^0 M}
}\Big)\quad.
\end{flalign}
The associated retarded/advanced Green's homotopies read as
\begin{flalign}
\parbox{10cm}{\xymatrix{
\ar[d]_-{\subseteq}\Omega^0_K(M) \ar[r]^-{\dd} 
~&~ \ar[dl]_-{\delta E^\pm}\ar[d]_-{\subseteq} \Omega^1_K(M) \ar[r]^-{\delta\dd} 
~&~\ar[dl]_-{E^\pm}\ar[d]_-{\subseteq}\ar[r]^-{\delta} \Omega^1_K(M) 
~&~\ar[dl]_-{\dd E^\pm}\ar[d]_-{\subseteq} \Omega^0_K(M)\\
\Omega^0_{J^\pm_M(K)}(M) \ar[r]_-{\dd} 
~&~ \Omega^1_{J^\pm_M(K)}(M) \ar[r]_-{\delta\dd} 
~&~\ar[r]_-{\delta} \Omega^1_{J^\pm_M(K)}(M) 
~&~\Omega^0_{J^\pm_M(K)}(M)
}}\quad,
\end{flalign}
where $E^\pm$ are the retarded/advanced Green's operators 
for the d'Alembertian $\Box = \delta\,\dd + \dd\,\delta$.
\end{ex}

The theory of Green's witnesses and Green's homotopies allows us to associate 
to any free BV theory on $\Loc_m$, endowed with a natural Green's witness, a dg-AQFT
\begin{flalign}\label{eqn:AQFTconstruction}
\xymatrix@C=3em{
\big(F_M, Q_M, (-,-)_M,W_M\big)_{M\in\Loc_m}\ar@{~>}[rr]^-{\text{CCR quantization}}~&&~ \AAA\in\dgAQFT(\Loc_m,\perp)^{\mathrm{ho}W}
}
\end{flalign}
that satisfies the homotopy time-slice axiom.
The latter can be constructed via canonical-commutation-relations (CCR) 
quantization, see \cite{FreeFAvsAQFT} for the details.
This construction is compatible in the following sense
with the dg-tPFA construction \eqref{eqn:tPFAconstruction}: The functor
\eqref{eqn:AQFTtotPFA} generalizes to the dg-setting, yielding a right Quillen functor
\begin{flalign}\label{eqn:compdg}
\Phi^\ast\,:\,\dgAQFT(\Loc_m,\perp)^{\mathrm{ho}W} ~\longrightarrow~\mathbf{dgtPFA}(\Loc_m)^{\mathrm{ho}W}\quad.
\end{flalign}
Applying this functor to the dg-AQFT \eqref{eqn:AQFTconstruction} yields a dg-tPFA
that is isomorphic to \eqref{eqn:tPFAconstruction},
i.e.\ $\Phi^\ast(\AAA)\cong \FFF$ for any free BV theory on $\Loc_m$ with a
natural Green's witness.
\begin{open}
We expect, but currently do not know how to prove, that \eqref{eqn:compdg} induces 
a Quillen equivalence when restricted to suitably defined additive objects. A formal model categorical
proof of this claim, which would generalize the Comparison Theorem \ref{theo:comparison} from 
$1$-categorical targets $\TT$ to the richer dg-context, is an interesting and important
problem for future works.
\end{open}
\begin{rem}
For the construction of interacting
models via perturbative AQFT \cite{FR1,FR2} 
we refer the reader to the relevant chapters
of this volume or to the monograph \cite{Rejzner}.
\end{rem}

%%%%%%%%%%%%%%%%%%%%%%%%%%%%%%%%%%%%%%%%%%%%%%%%
%%%%%%%%%%%%%%%%%%%%%%%%%%%%%%%%%%%%%%%%%%%%%%%%

\section{\label{sec:higher}Higher categories}

\paragraph{Why cochain complexes aren't enough:} The unital associative
(dg-)algebra that is assigned by a (dg-)AQFT $\AAA$ to a spacetime $M$ should be interpreted as
a quantization of the function algebra on the moduli ``space'' of fields on $M$, i.e.\
\begin{flalign}\label{eqn:AAAMassignment}
\AAA(M)\,=\, \mathcal{O}_\hbar\Big(\text{ moduli ``space'' of fields on $M$ }\Big)\,\in\,\Alg_{\mathsf{uAs}}(\Ch_\bbK)\quad.
\end{flalign}
It therefore depends strongly on the type of ``spaces'' one considers whether or not such an assignment is sensible.
For instance, in the context of (derived) algebraic geometry, there is just a small class of spaces, called
(derived) affine schemes, that are captured by their function algebras. Other spaces, such as (derived) 
stacks, are in general not. For example, considering the classifying 
stack $\mathbf{B}G:=[\ast/G]$ associated with a reductive 
affine group scheme $G$, one finds that its function algebra
\begin{flalign}
\mathcal{O}(\mathbf{B}G) \,\simeq \,N^\bullet(G,\bbK)\,\simeq\,\bbK\,=\,\mathcal{O}(\ast)
\end{flalign}
is weakly equivalent to that of a point, i.e.\ all information about the group $G$ gets lost. (Here
$N^\bullet$ denotes normalized group cochains.) This insufficiency of function algebras 
is a non-perturbative gauge-theoretic feature. Indeed, considering only the formal neighborhood 
of a point $x:\ast \to X$ in a (derived) stack $X$,
i.e.\ working perturbatively around a given field, one obtains a formal moduli problem
that admits an algebraic description in terms of $L_\infty$-algebras or dually in terms of their 
Chevalley-Eilenberg dg-algebras, see \cite{LurieFMP,Pridham}.
\sk

This issue is well-known to (derived) algebraic geometers,
who have proposed an interesting solution, see e.g.\ \cite{Toen} for an informal overview
and \cite{CPTVV} for the technical details. Instead of assigning to a (derived) stack
$X$ its function algebra $\mathcal{O}(X)\in \Alg_{\mathbb{E}_\infty}(\Ch_\bbK)$,
which as we have seen above is insufficient, one should better consider its symmetric
monoidal  dg-category of quasi-coherent modules $\mathbf{QCoh}(X)\in\Alg_{\mathbb{E}_\infty}(\dgCat_\bbK)$.
Informally speaking, the latter describes the (dg-)vector bundles over $X$, hence it is a priori richer
than the function algebra, which one can think of as the sections of the trivial line bundle.
As evidence that this approach is better behaved, let us consider again the classifying
stack $\mathbf{B}G$ associated with a reductive affine group scheme. One can show 
(see e.g.\ \cite[Proposition 2.17]{CotangentStack} for a spelled out proof) that the associated
symmetric monoidal dg-category of quasi-coherent modules 
\begin{flalign}
\mathbf{QCoh}(\mathbf{B}G)\,\simeq\, \Ch_\bbK^{\mathcal{O}(G)}\,=:\,\mathbf{dgRep}(G)
\end{flalign}
is given by the symmetric monoidal dg-category of $\O(G)$-dg-comodules, or in 
other words $G$-representations on cochain complexes.
The group $G$ can be reconstructed from this symmetric monoidal 
dg-category via a Tannakian reconstruction argument, hence
no information is lost.
\sk

Transferring these ideas to the context of AQFT motivates us to replace \eqref{eqn:AAAMassignment} 
with the assignment of quantized dg-categories (in the sense of \cite{Toen,CPTVV}) to spacetimes $M$,
which we informally interpret as
\begin{flalign}
\AAAA(M)\,=\,\mathbf{QCoh}_{\hbar}\Big(\text{ moduli ``space'' of fields on $M$ }\Big)\,\in\dgCat_\bbK\quad.
\end{flalign}
In the special case where the moduli ``spaces'' of fields are only $1$-stacks, without derived structure,
one can simplify this framework by considering (locally presentable) $\bbK$-linear categories 
instead of dg-categories. In the running example
given by the classifying stack $\mathbf{B}G$, this simplification yields a restriction from dg-representations 
$\mathbf{dgRep}(G)$ to the ordinary symmetric monoidal category $\mathbf{Rep}(G) := \Vec_{\bbK}^{\mathcal{O}(G)}$
of $\mathcal{O}(G)$-comodules. In the paragraphs below, we shall start with this simplified setting
because it can be formalized in the simpler context of $2$-category theory instead of $\infty$-category theory. 

\paragraph{Foundations for $2$-AQFTs:} Let us denote by $\Pr_\bbK$ the closed symmetric monoidal
$2$-category of locally presentable $\bbK$-linear categories, co-continuous $\bbK$-linear functors 
and natural transformations. See \cite{2AQFT} for a concise review. 
$\Pr_\bbK$ serves as the target for the $2$-categorical generalization
of AQFT that implements the ideas explained in the previous paragraph.
\begin{defi}\label{def:2AQFT} 
The $2$-category of \textit{$2$-AQFTs} on an orthogonal category $(\CC,\perp)$ is defined as the $2$-category
of weak algebras over the operad $\mathcal{P}_{(\CC,\perp)}$ 
(see \eqref{eqn:AQFToperadspecial} for its relation to the AQFT operad) 
with values in $\mathbf{Pr}_\bbK$, i.e.\
\begin{flalign}
\mathbf{2AQFT}(\CC,\perp)\,:=\,\Alg^{\mathrm{weak}}_{\mathcal{P}_{(\CC,\perp)}}\big(\mathbf{Pr}_\bbK\big)\quad.
\end{flalign}
For the technical details see \cite{2AQFT}.
\end{defi}
\begin{rem}
Spelling this out in more detail, a $2$-AQFT $\AAAA\in\mathbf{2AQFT}(\CC,\perp)$
is an assignment
\begin{subequations}\label{eqn:2AQFTpictures}
\begin{flalign}\label{eqn:2AQFTpictures1}
\parbox{8cm}{
\begin{tikzpicture}[scale=0.5]
\filldraw[fill=gray!20, draw=black] (0,0) -- (1.8,1.8) -- (3.6,0) -- (1.8,-1.8) -- (0,0);
\draw (3.1,1.3) node {\footnotesize{$M$}};
\draw[very thick, |->] (5,0) -- (8,0) node[midway,above] {{\footnotesize{\text{$2$-AQFT}}}};
\draw (11.5,0) node {$\AAAA(M) \in \Pr_{\bbK}$};
\end{tikzpicture}}
\end{flalign}
of a locally presentable $\bbK$-linear category to each object $M\in \CC$, 
together with the assignment
\begin{flalign}\label{eqn:2AQFTpictures2}
\parbox{8cm}{
\begin{tikzpicture}[scale=0.5]
\filldraw[fill=gray!20, draw=black] (0,-4) -- (1.8,-2.2) -- (3.6,-4) -- (1.8,-5.8) -- (0,-4);
\draw (3.1,-2.7) node {\footnotesize{$N$}};
\draw[thick, dotted] (1,-3.2) -- (2.6,-4.8);
\draw[thick, dotted] (1,-4.8) -- (2.6,-3.2);
\draw[fill=gray!70, draw=black] (0.2,-4) -- (0.7,-3.5) -- (1.2,-4) -- (0.7,-4.5) -- (0.2,-4);
\draw (0.7,-4) node {\tiny{$M_1$}};
\draw (1.8,-4) node {\text{{\bf $\cdots$}}};
\draw[fill=gray!70, draw=black] (2.4,-4) -- (2.9,-3.5) -- (3.4,-4) -- (2.9,-4.5) -- (2.4,-4);
\draw (2.9,-4) node {\tiny{$M_n$}};
\draw[very thick, |->] (5,-4) -- (8,-4) node[midway,above] {{\footnotesize{\text{$2$-AQFT}}}};
\draw (13,-4) node {$\bigboxtimes\limits_{i=1}^n \AAAA(M_i) ~\longrightarrow ~\AAAA(N)$};
\end{tikzpicture}}
\end{flalign}
\end{subequations}
of a co-continuous $\bbK$-linear functor to each tuple $\big(f_i:M_i\to N\big)_{i=1,\dots,n}$ of mutually
orthogonal $\CC$-morphisms, i.e.\ $f_i\perp f_j$ for all $i\neq j$, 
and of coherence data witnessing the compatibility
with compositions, units and permutation actions.
Comparing this with ordinary AQFTs \eqref{eqn:AQFTpictures}, we note that
$2$-AQFTs are simply given by replacing the category of unital associative algebras
$\Alg_{\mathsf{uAs}}(\Vec_\bbK)$ by the $2$-category $\mathbf{Pr}_\bbK$ of 
locally presentable $\bbK$-linear categories.
This is precisely the generalization that was motivated and explained in the previous paragraph.
\end{rem}

\begin{rem}
Higher algebraic structures that are similar to the $2$-AQFTs from Definition \ref{def:2AQFT} 
also appeared in the study of category valued factorization homology
by Ben-Zvi, Brochier and Jordan \cite{BZBJ,BZBJ2}.
\end{rem}

The most pressing question is of course whether or not the concept of $2$-AQFTs
is indeed a generalization of the ordinary concept of AQFTs from Section \ref{sec:operad}.
The following result, which was proven in \cite[Theorem 4.3]{2AQFT}, shows that
this is the case.
\begin{theo}\label{theo:2adjunction}
For each orthogonal category $(\CC,\perp)$, there exists a bicategorical adjunction
\begin{flalign}\label{eqn:incltrunc}
\xymatrix{
\iota\,:\,\AQFT(\CC,\perp)\ar@<0.75ex>[r]~&~\ar@<0.7ex>[l] \mathbf{2AQFT}(\CC,\perp)\,:\,\pi
}
\end{flalign}
that exhibits ordinary AQFTs as a coreflective full $2$-subcategory of the $2$-category
of $2$-AQFTs.
\end{theo}

In simple terms, this result states that ordinary AQFTs can be considered equivalently
as special examples of $2$-AQFTs by applying the inclusion functor $\iota$. 
The $2$-AQFT $\iota(\AAA)$ associated with an ordinary 
$\AAA\in \AQFT(\CC,\perp)$ takes a particularly simple
form: It assigns to $M\in\CC$ the category $\iota(\AAA)(M) = {}_{\AAA(M)}\mathbf{Mod}\in\mathbf{Pr}_\bbK$ 
of modules over the algebra $\AAA(M)\in \Alg_{\mathsf{uAs}}(\Vec_\bbK)$
and to a tuple of mutually orthogonal $\CC$-morphisms the induced module functor 
associated with the algebra map $\bigotimes_{i=1}^n\AAA(M_i)\to \AAA(N)$.
\sk

It is natural to ask whether there exist $2$-AQFTs that do 
not arise from ordinary AQFTs via the inclusion functor $\iota$. We call
such $2$-AQFTs \textit{non-truncated}, while the ones arising from $\iota$ 
are called \textit{truncated}. The mathematical
problem can be formulated as follows: Find objects $\AAAA\in \mathbf{2AQFT}(\CC,\perp)$ such that 
the counit $\epsilon_{\AAAA} : \iota\pi(\AAAA)\to \AAAA$ of the adjunction \eqref{eqn:incltrunc}
is not an equivalence. Simple examples of non-truncated $2$-AQFTs can be obtained
by an orbifold construction: Suppose that $\AAA\in \AQFT(\CC,\perp)$ is an ordinary
AQFT that is endowed with the action of a finite group $G$. The
associated orbifold $2$-AQFT $\AAA^G\in \mathbf{2AQFT}(\CC,\perp)$
is defined by assigning to $M\in\CC$ the category
\begin{flalign}
\AAA^G(M)\,:=\, {}_{\AAA(M)}\mathbf{Mod}^{\mathcal{O}(G)}\,\in\,\mathbf{Pr}_\bbK
\end{flalign}
of $G$-equivariant $\AAA(M)$-modules and to a tuple of mutually orthogonal $\CC$-morphisms 
the associated induced module functor, see \cite[Section 5]{2AQFT} for the details.
The question whether or not the  $2$-AQFT $\AAA^G$ is truncated 
can be reduced to an algebraic problem, see \cite[Theorem 5.11]{2AQFT}.
\begin{theo}
$\AAA^G\in \mathbf{2AQFT}(\CC,\perp)$ is truncated if and only if
the algebra extension $\AAA(M)^{\mathcal{O}(G)}\subseteq \AAA(M)$ of $G$-invariants 
is $\mathcal{O}(G)$-Hopf-Galois, for all $M\in \CC$.
\end{theo}
\begin{ex}
The simplest example for a non-truncated $2$-AQFT is given by the orbifold
construction $\bbK^G\in \mathbf{2AQFT}(\CC,\perp)$ associated with the trivial AQFT $\bbK\in \AQFT(\CC,\perp)$
and the trivial action of a non-trivial finite group $G$. 
\end{ex}

\paragraph{Towards $\infty$-AQFTs:} The previous paragraph should be seen
as a warm-up for the richer case where the target category
is the symmetric monoidal $\infty$-category $\dgCat_\bbK$ of dg-categories.
This will cover a whole range of new examples arising from
quantizations of derived stacks. 
\begin{defi}\label{defi:inftyAQFT}
The $\infty$-category of \textit{$\infty$-AQFTs} on an orthogonal category $(\CC,\perp)$
is defined as the $\infty$-category of algebras 
over the operad $\mathcal{P}_{(\CC,\perp)}$ (in the sense of \cite{LurieHA})
with values in $\dgCat_\bbK$, i.e.\
\begin{flalign}
\infty\AQFT(\CC,\perp)\,:=\,\Alg_{\mathcal{P}_{(\CC,\perp)}}^{\infty}\big(\dgCat_\bbK\big)\quad.
\end{flalign}
\end{defi}

Unfortunately, this framework is presently not explored in much detail. Let us however note that there is
a construction that assigns to any dg-AQFT
$\AAA\in\dgAQFT(\CC,\perp)$ an $\infty$-AQFT. The latter is given by assigning
to objects $M\in \CC$ the dg-categories ${}_{\AAA(M)}\mathbf{dgMod}\in \dgCat_\bbK$
of $\AAA(M)$-dg-modules and to tuples of mutually disjoint $\CC$-morphism
the induced dg-module functor. 
\begin{open}
Generalize Theorem \ref{theo:2adjunction} by establishing an $\infty$-adjunction
\begin{flalign}
\xymatrix{
\iota\,:\,\dgAQFT(\CC,\perp)\ar@<0.75ex>[r]~&~\ar@<0.7ex>[l] \infty\AQFT(\CC,\perp)\,:\,\pi
}
\end{flalign}
that relates the dg-AQFTs from Section \ref{sec:homotopy} to $\infty$-AQFTs. It would be particularly
interesting to prove that this $\infty$-adjunction exhibits the $\infty$-category associated with the model
category $\dgAQFT(\CC,\perp)$ as a coreflective full $\infty$-subcategory $\infty\AQFT(\CC,\perp)$.
\end{open}

Despite the currently undeveloped foundations of $\infty$-AQFTs,
there is promising evidence that interesting examples for such objects
can be obtained from quantization constructions in derived algebraic geometry
\cite{CPTVV,Toen}. A first attempt towards constructing examples 
was taken in \cite[Section 4]{CotangentStack}. In more detail,
consider the canonical phase space of Yang-Mills theory with 
structure group any reductive affine group scheme $G$
on a lattice approximation of the Cauchy surface $\Sigma$. 
This phase space is the derived cotangent stack
\begin{flalign}
T^\ast\big[\mathrm{Con}_G(\Sigma)\big/\mathcal{G}(\Sigma)\big]
\end{flalign}
over the quotient stack of $G$-connections modulo the gauge group on $\Sigma$,
together with the canonical unshifted symplectic structure. A quantization
of the dg-category of quasi-coherent modules over this derived stack
was described in \cite[Proposition 3.13]{CotangentStack}, which can be interpreted
in terms of $D$-modules over the quotient stack $\big[\mathrm{Con}_G(\Sigma)\big/\mathcal{G}(\Sigma)\big]$.
It was then shown in \cite[Proposition 4.10]{CotangentStack} that the assignment
\begin{flalign}
\AAAA(\Sigma)\,:=\, \mathbf{QCoh}_{\hbar}\big(T^\ast\big[\mathrm{Con}_G(\Sigma)\big/\mathcal{G}(\Sigma)\big]\big)\,\in\dgCat_{\bbK}
\end{flalign}
of these quantized dg-categories to lattices (or more generally to directed graphs) can be promoted to
a (lattice variant of) an $\infty$-AQFT in the sense of Definition \ref{defi:inftyAQFT}.
\begin{open}
It would be interesting to construct examples of $\infty$-AQFTs on \textit{spacetime} 
lattices instead of spatial ones.
This would allow one to investigate the time-slice axiom
in the context of $\infty$-AQFT, which tentatively should be of the form
that each Cauchy morphism $f: M\to N$ is assigned
to a quasi-equivalence $\AAAA(f): \AAAA(M)\stackrel{\sim}{\to}\AAAA(N)$ of dg-categories.
\end{open}

%%%%%%%%%%%%%%%%%%%%%%%%%%%%%%%%%%%%%%%%%%%%%%%%
%%%%%%%%%%%%%%%%%%%%%%%%%%%%%%%%%%%%%%%%%%%%%%%%
\section*{Acknowledgments}
The work of M.B.\ is fostered by 
the National Group of Mathematical Physics (GNFM-INdAM (IT)). 
A.S.\ gratefully acknowledges the support of the Royal Society (UK) through a Royal Society University
Research Fellowship (URF\textbackslash R\textbackslash 211015) and an 
Enhancement Grant (RF\textbackslash ERE\textbackslash 210053). We also would like
to thank our collaborators who have contributed and/or currently contribute 
to the homotopical AQFT research program:
A.~Anastopoulos, S.~Bruinsma, S.~Bunk, V.~Carmona, C.~J.~Fewster, 
L.~Giorgetti, A.~Grant-Stuart, J.~MacManus, G.~Musante, M.~Perin, J.~P.~Pridham, P.~Safronov, U.~Schreiber, 
R.~J.~Szabo and L.~Woike.

%%%%%%%%%%%%%%%%%%%%%%%%%%%%%%%%%%%%%%%%%%%%%%%%
%%%%%%%%%%%%%%%%%%%%%%%%%%%%%%%%%%%%%%%%%%%%%%%%

%%%%%%%%%%%%%%%%%%%%%%%%

\end{document}